\documentclass[aps,prd,nofootinbib,superscriptaddress
,notitlepage, showkeys, 11pt]{revtex4-1}
\usepackage{amsfonts} 
\usepackage{amssymb}  
\usepackage{graphicx} 
\usepackage{epsfig}  
\usepackage{amsmath}  
\usepackage{mathtools}
\begin{document}

\title{The photon time delay in magnetized vacuum magnetosphere}

\author{A.W. Romero Jorge}
\affiliation{Facultad de Física, Universidad de la Habana, San L\'azaro y L, Vedado,  La Habana, 10400, Cuba}
\affiliation{Departamento de Física Teórica, Instituto de Cibern\'{e}tica Matem\'{a}tica y F\'{\i}sica (ICIMAF), \\
Calle E esq 15 No. 309 Vedado, La Habana, 10400, Cuba}

\author{E. Rodr\'iguez Querts}
\affiliation{Departamento de Física Teórica, Instituto de Cibern\'{e}tica Matem\'{a}tica y F\'{\i}sica (ICIMAF), \\
Calle E esq 15 No. 309 Vedado, La Habana, 10400, Cuba}

\author{H. P\'erez Rojas }
\affiliation{Departamento de Física Teórica, Instituto de Cibern\'{e}tica Matem\'{a}tica y F\'{\i}sica (ICIMAF), \\
Calle E esq 15 No. 309 Vedado, La Habana, 10400, Cuba}

\author{A. P\'erez Mart\'inez}
\affiliation{Departamento de Física Teórica, Instituto de Cibern\'{e}tica Matem\'{a}tica y F\'{\i}sica (ICIMAF), \\
Calle E esq 15 No. 309 Vedado, La Habana, 10400, Cuba}

\author{L. Cruz Rodr\'iguez}
\affiliation{Facultad de Física, Universidad de la Habana, San L\'azaro y L, Vedado,  La Habana, 10400, Cuba}

\author{G. Piccinelli Bocchi}
\affiliation{Centro Tecnol\'ogico FES Arag\'on, Universidad Nacional Aut\'onoma de M\'exico, Avenida Rancho Seco S/N, Bosques de Arag\'on, 57130 Nezahualc\'oyotl, Estado de M\'exico, M\'exico}

\author{J. A. Rueda}
\affiliation{ICRANet, Piazza della Repubblica 10, I–65122 Pescara, Italy}
\affiliation{ICRANet-Ferrara, Universit\`a degli Studi di Ferrara, Via Saragat 1, I-44122 Ferrara, Italy}
\affiliation{Dipartimento di Fisica e Scienze della Terra, Universit\`a degli Studi di Ferrara, Via Saragat 1, I-44122 Ferrara, Italy}
\affiliation{Istituto di Astrofisica e Planetologia Spaziali, INAF, Via Fosso del Cavaliere 100, I-00133 Rome, Italy}

\begin{abstract}
We study the transverse propagation of photons in a magnetized vacuum considering radiative corrections in the one-loop approximation. The dispersion equation is modified due to the magnetized photon self-energy in the transparency region ($0<\omega<2m_e$). The aim of our study is to explore propagation of photons in a neutron star magnetosphere (described by magnetized vacuum). The solution of the dispersion equation is obtained in terms of analytic functions.
The larger the magnetic field, the lower the phase velocity and the more the dispersion curve deviates from the light-cone.
For fixed values of the frequency, we study the dependence of photon time delay with the magnetic field strength, as well as with distance. For the latter, we adopt a magnetic dipole configuration and obtain that -contrary to the expectation, according to the traditional time delay of photons in the interstellar medium- photons of higher energy experience a longer time delay. A discussion of potential causes of this behaviour is presented.
\end{abstract}
\maketitle

\section{Introduction}\label{sec1}

Neutron stars are mysterious compact objects where strong gravitational and electromagnetic fields emerge. These objects usually manifest as pulsars, emitting energetic electromagnetic signals detected at very precise intervals \citep{Camenzind}.

Even though they were discovered more than fifty years ago, many ingredients of these astrophysical systems are still poorly known: the composition of the compact object itself \citep{Weber2005}, the composition and structure of their magnetosphere \citep{Petri2016}, and the generation of their strong magnetic fields \citep{Duncan1992, Dieters1998}.

Concerning the magnetosphere, three fundamental models can be considered: the simplest one consists of a naked star, without any kind of plasma in its neighborhood; the opposite one in which we have the compact object fully immersed in a plasma, and finally, an intermediate model that admits the existence of a magnetosphere partially filled with electrons and positrons \citep{Petri2016}

On the other hand, a magnetic field strength on the surface of neutron stars of the order of Schwinger’s critical field $B_c= 4.41 \times 10^{13}$~G (e.g. \citep{Ciolfi2014}), and beyond ( e.g. \citep{Dieters1998}), imposes to study the behavior of matter at extreme conditions.

On the observational side, the information we can get from pulsars is based on the measurement of their pulses arrival times. During their propagation from the source to Earth, photons can experience a variety of time delays. In particular, a most known time delay is the dispersion of photons when interacting with the electrons in the interstellar medium \citep{Pushkarev2010},\citep{Wang2004}, \citep{Bosnjakl2012}, \citep{Waxman1996}. This type of time delay mainly depends on the electron column density and the distance to the source, and causes that pulses at lower frequencies are delayed with respect to those emitted at higher frequencies (the delay is inversely proportional to the square of the photon frequency; see e.g.~\citep{2004hpa..book.....L}).

Therefore, the knowledge of this type of time delay probes the interstellar medium properties rather than the source site ones.

We are here interested in a photon time delay process occurring in the vicinity of the source, and unrelated with the photon interaction with matter. The propagation of light in vacuum is modified by various external agents: electromagnetic fields, temperature, geometric boundary configurations, gravitational background and non-trivial topologies. In particular, the problem of light propagation in electron-positron vacuum in the presence of a magnetic field is similar to the dispersion of light in an anisotropic medium, where the external field axis sets the anisotropic direction. Therefore, the photon dispersion relation is corrected by adding the polarization tensor $\Pi(k_{\perp},k_{\parallel}, B,\omega)$. This takes into account the indirect interaction with the magnetic field through the virtual electron-positron pairs \citep{PerezRojasH.Shabad1978} and depend on components of the wave vector, the photon frequency and the external magnetic field.

In this paper, we study photon time delay which might occur in the pulsar magnetosphere. We use the simplest approximation to describe it, which consists in considering it as a magnetized vacuum.

We start by solving the photon dispersion equation considering the radiative corrections given by the magnetized photon self-energy. Then, we compute the phase velocity and the photon time delay.

The propagation of photons is considered perpendicular to the magnetic field ($k \perp  B)$ since, for parallel propagation, photons behave like in absence of the magnetic field, namely with no deviation from the light-cone.

From a physical aspect, we estimate the time delay of photons in the region of the pulsar magnetosphere, modeling it as an electromagnetic vacuum.
From a mathematical point of view, our calculation is more robust (and elegant) than others \citep{PerezRojas2014}, since our analytic expressions for the solution of the dispersion equation are presented in term of A-hypergeometric functions \citep{Sturmfels2000}.

The paper is organized as follows. In section \ref{sec2}, we solve the dispersion equations considering the radiative corrections given by the photon self-energy in presence of magnetic field. We devote section \ref{sec3} to discuss the phase velocity and time delay of photons traveling in a magnetized vacuum. The dependence with the radial coordinate of the photon delay is studied assuming a dipole configuration for the magnetic field in the magnetosphere. Finally, we present in section~\ref{sec4} the conclusions of our work.

\section{Propagation of photon in magnetized vacuum}  \label{sec2}

In this section, we study the propagation of photons perpendicular to the constant and uniform external magnetic field in  vacuum\footnote{We use natural units $\hbar=c=1$.}.

It is well known that photons in  vacuum obey the dispersion equation
\begin{equation}\label{lightcone}
k^2_{\perp}+k^2_{\parallel}-\omega^2=0,
\end{equation}
that implies that photons travel at the speed of light.

The effect of the presence of the magnetic field on the dispersion relation, Eq.~(\ref{lightcone}), can be included through radiative corrections to the photon self-energy.
The modified dispersion equations ~\citep{Sha1984} are
\begin{equation}
	k^2=\kappa^{(i)}(\omega,k_{\parallel},k_{\perp},b),
\end{equation}
where $b$ is the magnetic field normalized to the Schwinger's field, $b=B/B_c$, and $\kappa^{(i)}$ are the eigenvalues of the photon self-energy given in the appendix.

In what follows, we consider photon propagation perpendicular to the magnetic field  ($\vec{k}\perp \vec{B}$). Three modes appear: one longitudinal mode $i=1$, that is not physical and two transverse ones $i=2,3$. The threshold of pair creation of second and third modes are $\omega=2m_e$ and  $\omega=m_e+\sqrt{m_e^2+2eB}$, respectively \citep{PerezRojas2014}.

In our study, we consider only the second mode, which is more relevant in the region of transparency.  The corrections of dispersion relations become relevant close to the thresholds, and the  second mode threshold  is independent on the magnetic field, being  much lower than the threshold for the third mode, for the considered values of the magnetic field.

For a large range of frequencies, the solution of the dispersion equation corresponds to relatively small deviations from the light-cone, $k^2\ll e B$, except for values of $\omega^2-k^2_{\parallel}$ extremely close to the vacuum threshold for pair creation \citep{PerezRojas2014}.  As it is   shown in the appendix, in this case, we can write the photon self-energy eigenvalues as polynomials in $k^2$:
\begin{equation}
    \kappa_{i}=\sum_{l=0}^{\infty}\chi_{il}(k^2)^l.
\end{equation}

If we truncate the first four terms of the power series, for $j=3$ we obtain a cubic equation in $k^2$. This equation has been solved in \citet{PerezRojas2014} using Cardano formulas for polynomials of third degree. However, numerical calculations with quadratic and cubic roots are thorny, so in this work we solve it with the aid of hypergeometric functions \citep{Sturmfels2000}:
	\begin{equation}
	k^2=-\sum_{j_2,...,j_n=0}^{\infty}\dfrac{(-1)^{j_1}j_1!}{(j_0+1)!j_2!...j_n!}\dfrac{\chi_{0}^{j_0+1} \chi_{i2}^{j_2} ...\chi_{in}^{j_n} }{(\chi_{i1}-1)^{j_1+1}},
	\end{equation}
where $ j_0=j_2+2j_3+...+(n-1)j_n $, $j_1=2j_2+3j_3+...+nj_n $, and integral expressions of $\chi_{i1}$ are written in the appendix.

\begin{figure}[t]
		\centering
\includegraphics[width=.6\linewidth]{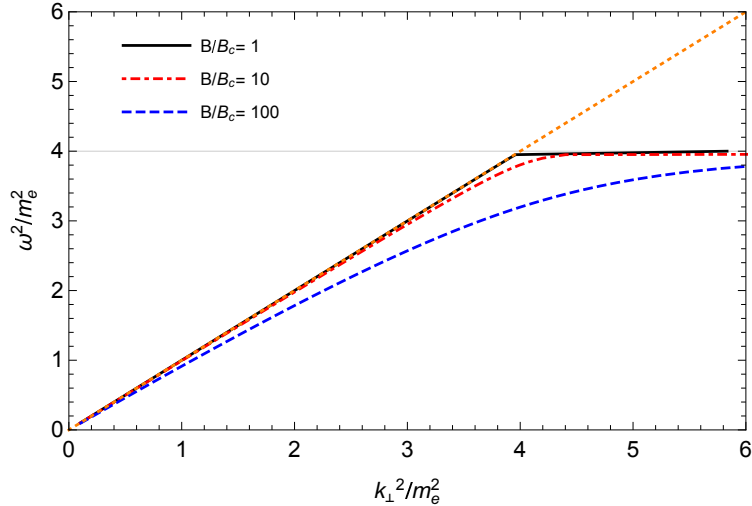}
		\caption{Dispersion relation for selected values of the magnetic field strength. The orange line corresponds to the propagation of light for $B=0$ (light-cone) and the gray line marks the first threshold of pair creation. We recall that $B_{c}=4.41 \times 10^{13}$~G and $m_{e}= 0.511$~MeV.}
		\label{fig:lightcone}
\end{figure}

The solution of the dispersion equation is shown in Fig.~\ref{fig:lightcone} for selected values of the magnetic field strength. The figure shows that, when the magnetic field increases, the deviation from the light-cone is higher. Besides, for any value of the magnetic field, a threshold exists $\omega=2m_{e}$, above which the photons have a high probability to decay in electron-positron pairs \citep{PerezRojas2009}.

For our purposes, we are only interested in the study of the region of transparency ($ 0<\omega<2m_e$), which is the region of momentum space where the photon self-energy and its frequency have real values.

\section{Phase velocity and time delay in magnetized vacuum} \label{sec3}

In this section, we calculate the phase velocity and the photon time delay taking advantage of the previous calculations.

The photon phase velocity takes the form
\begin{eqnarray}
	v_{ph}(\omega,B)&=&\dfrac{\omega}{k_{\perp}}\nonumber\\
	&=&\left (1-\dfrac{1}{\omega^2}\sum_{j_2,j_3=0}^{\infty}\dfrac{(-1)^{j_1}j_1!}{(j_0+1)!j_2!j_3!}\dfrac{\chi_{i0}^{j_0+1} \chi_{i2}^{j_2} \chi_{i3}^{j_3}}{(\chi_{i1}-1)^{j_1+1}}   \right )^{-1/2}.
\end{eqnarray}
	
\begin{figure}[h!]
	\centering
		\includegraphics[width=.6\linewidth]{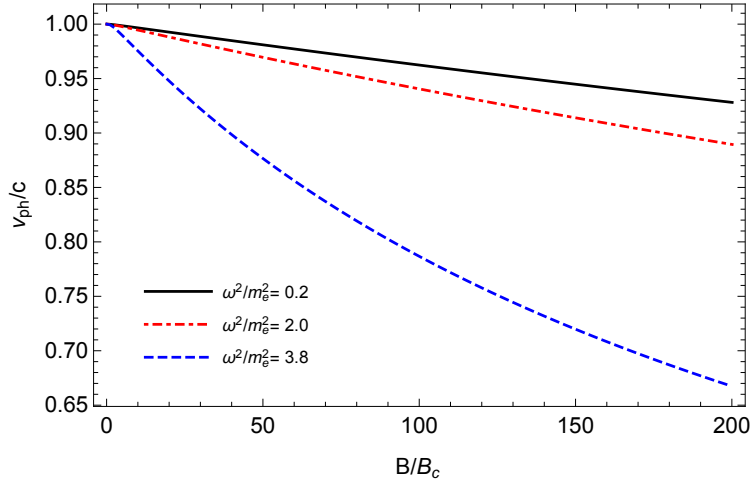}
		\caption{ Phase velocity in function of the  magnetic field  for different frequencies }
		\label{figure2}
\end{figure}

Figure~\ref{figure2} shows the photon phase velocity as a function of magnetic field for fixed values of the frequencies.

We can see that photons of higher energies have lower phase velocity than the lower energy ones, hence the latter suffer a longer time delay. Besides, we can appreciate that in the limit of low frequency (black solid line), the phase velocity decreases linearly with the external magnetic field strength.

\subsection{Photon  time delay in magnetosphere}

To calculate the photons time delay when crossing the magnetosphere (magnetized vacuum), for different energies, we consider a magnetic dipole configuration:
\begin{equation}\label{dipolar}
B(r)=B_{0} \left (\dfrac{r_{0}}{r}\right )^3,
\end{equation}
where $B_{0}$ and $r_{0}$ are, respectively, the surface magnetic field and radius of the neutron star.

We consider for $B_{0}$ values from $10^{12}$~G all the way up to $10^{15}$~G, covering the range of (theoretically) estimated fields from radio pulsars to soft gamma repeaters and anomalous X-ray pulsars (``magnetars'') \citep[e.g.][]{Duncan1992,Dieters1998}.
	
Using Eq.~(\ref{dipolar}), we can compute the time delay of the radiation crossing the magnetosphere of the pulsars given by the expression
\begin{equation}
    \tau=\int_{r_o}^ r \frac{dr}{v_{ph}(\omega,B(r))}.
\end{equation}
	
\begin{figure}[h!]
		\centering
		\includegraphics[width=.6\linewidth]{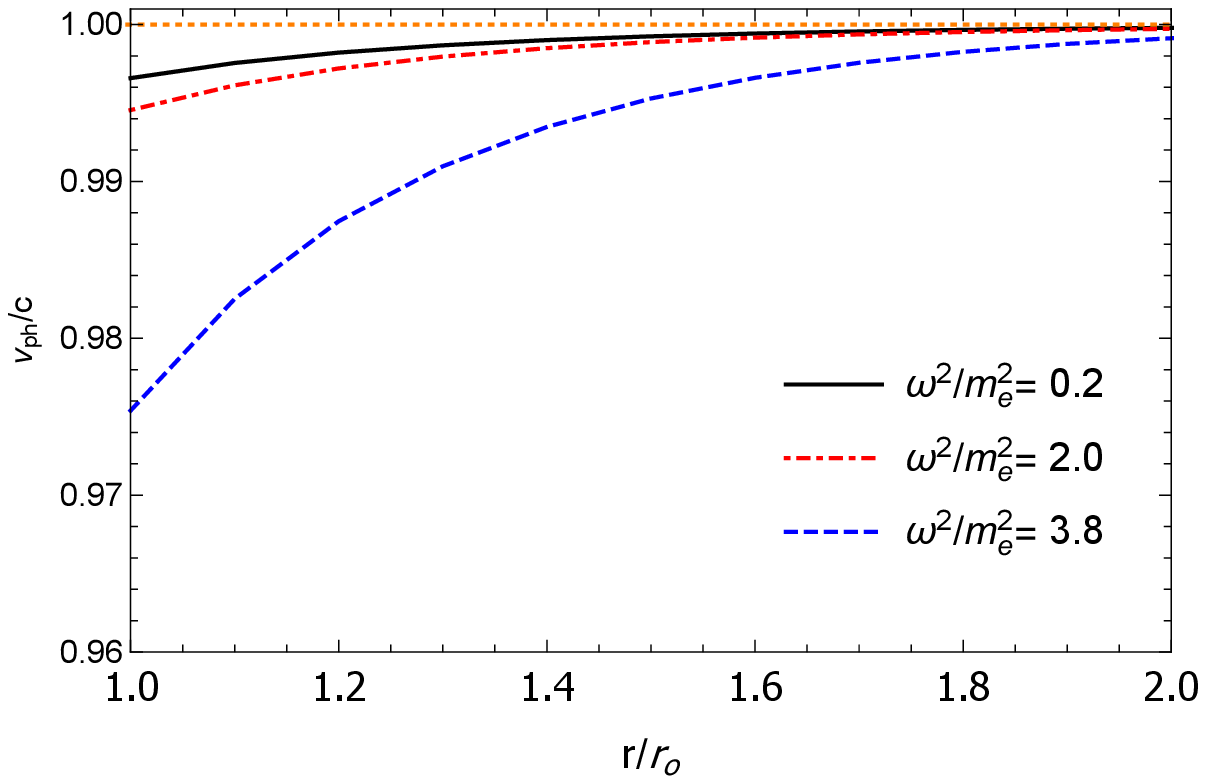}
		\includegraphics[width=.6\linewidth]{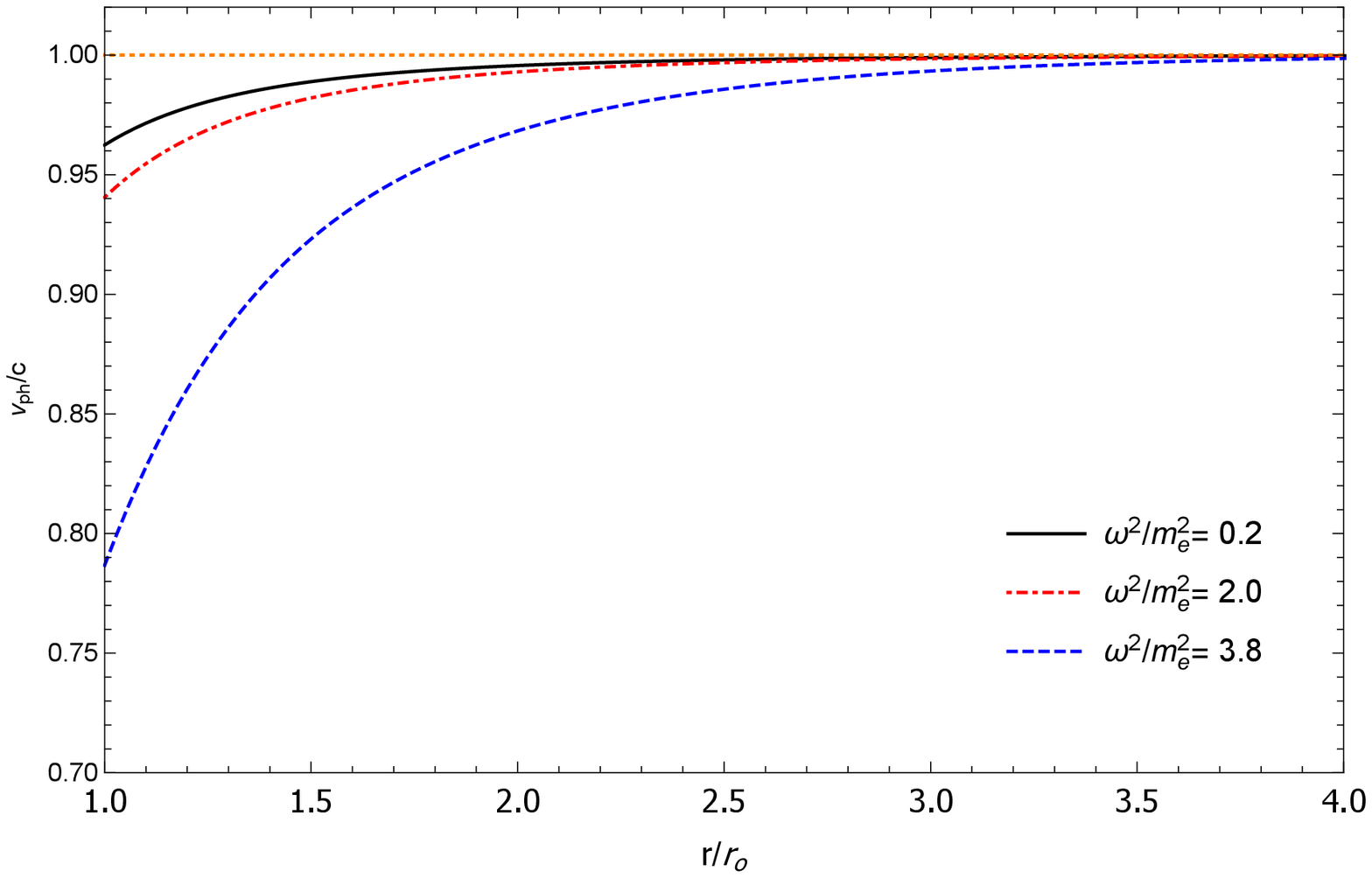}
		\caption{Phase velocity as a function of the distance traveled by the photons for different frequencies, top panel ($B_{0}= 10 B_{c}$) and bottom panel ($B_{0}= 100 B_{c}$) }
		\label{Velocidaddefase}
	\end{figure}

Figure~\ref{Velocidaddefase} shows the phase velocity as a function of the distance traveled by the photons for different values of frequencies and two different values of $B_{0}$. It can be seen how, as expected, the phase velocity tends to the speed of light as the magnetic field decreases.
	
\begin{figure}[h!]
		\centering
		\includegraphics[width=.6\linewidth]{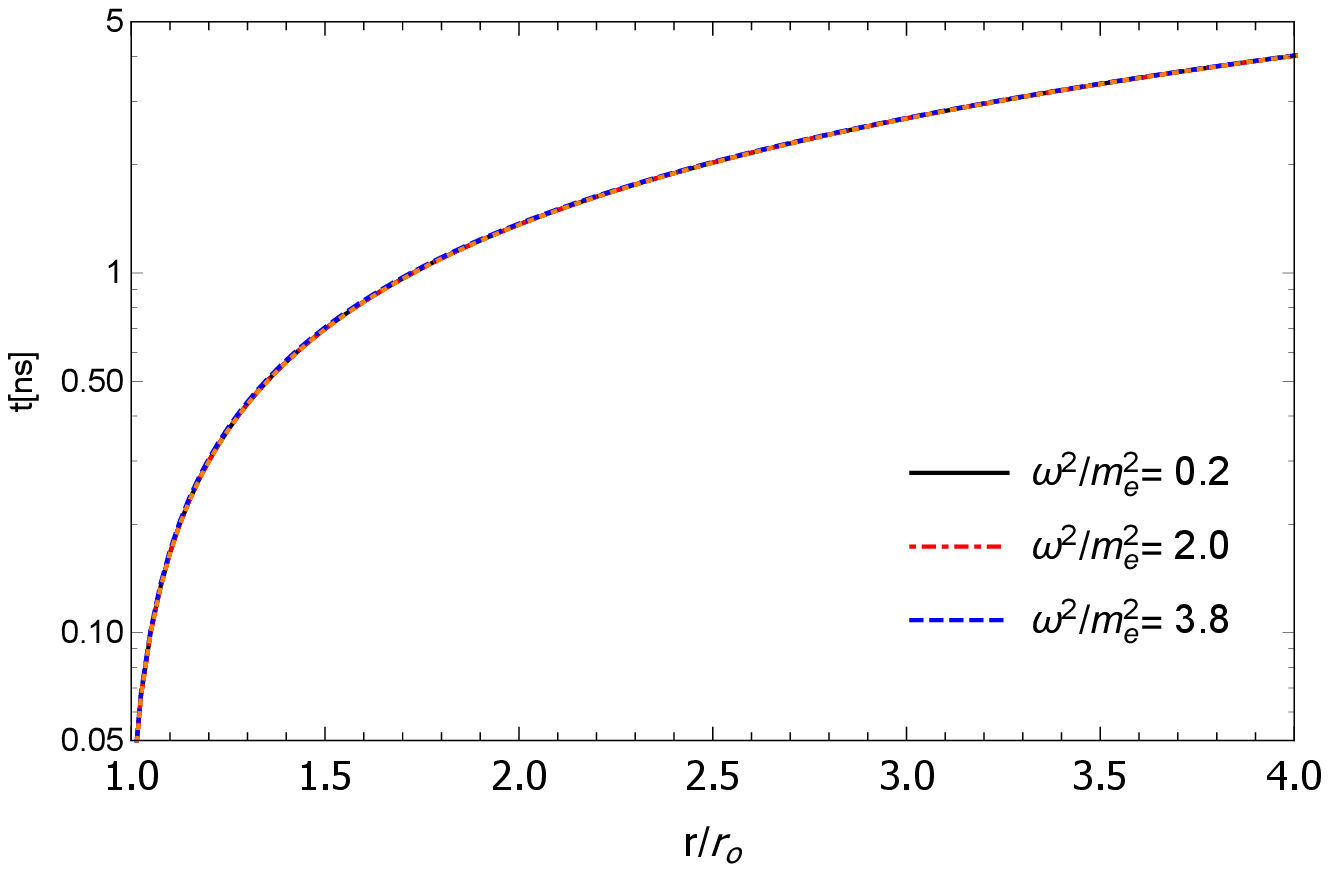}
			\includegraphics[width=.6\linewidth]{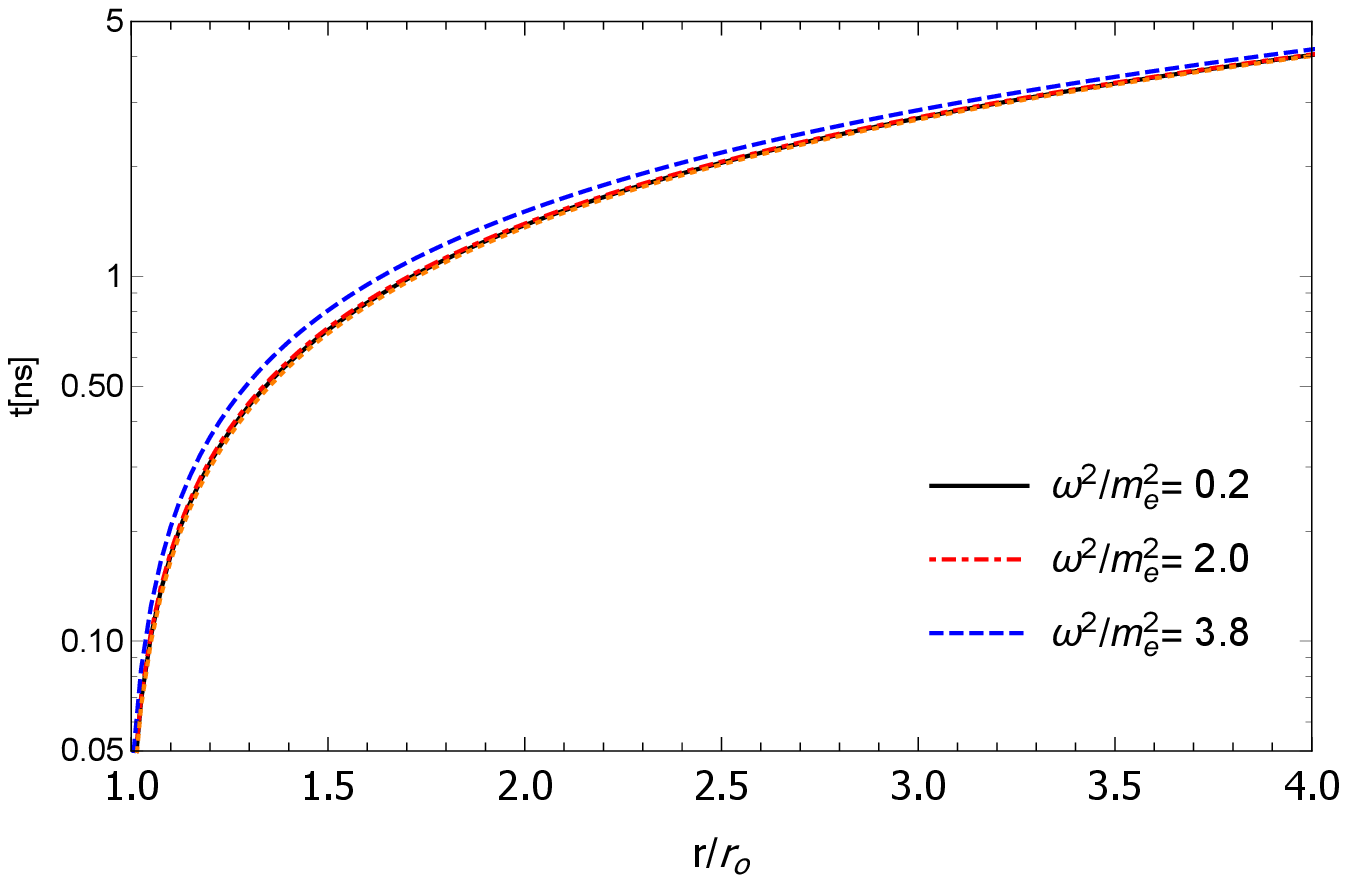}
		\caption{Time delay of photons for fixed values of  frequencies as a function of the distance traveled, top panel ($B_{o}= 10 B_{c}$) and bottom panel ($B_{o}= 100 B_{c}$) }
		\label{timedelay}
	\end{figure}

Figure~\ref{timedelay} shows the time delay of photons as a function of the distance, for two fixed values of the surface magnetic field $B_{0}$. The time delay is of the order of nanosecond for different values of frequencies. In spite of its shortness, it already shows that the solely presence of the magnetic field is sufficient to cause a time delay that grows with the photon frequency.

We would like to stress that more complex configurations of the magnetosphere, including for instance the magnetized electron-positron plasma, might have relevant effects. A study in this direction is currently in progress.

\section{Conclusions}\label{sec4}

We solved the dispersion equation for photons propagating perpendicular to a constant and uniform magnetic field.
	
The  analytical and approximate expressions for the phase velocity, valid in a wide range of the characteristic parameters, have been calculated. The photon phase velocity depends on the magnetic field and the photon energy.

In the limit of low frequencies, the phase velocity tends to have a linear behavior with the magnetic field.
	
The photon time delay was calculated starting from a simple model of the magnetic field configuration in the neutron star magnetosphere. We found that differences between the photon time delay of $\gamma$-radiation $\lesssim 1~ MeV$ is of the order of nanosecond. This difference might be due to the fact that more energetic photons interact stronger with virtual electron-positron pairs,  being closer to the threshold ($\omega=2 m_{e}$)
	
In this work, we have made a first attempt to study the photon time delay considering a simple model of the magnetosphere. An improvement of our study should include, for instance, a more realistic model of pulsar magnetosphere and the magnetized electron-positron plasma, as well as other possible geometrical configurations of the magnetic field that surrounds the neutron star.

A relevant result of our work is that, contrary to the traditional time delay of photons in the interstellar medium, in the present quantum electrodynamical process the more energetic photons are delayed with respect to the lower energetic ones. An essential pending task is to understand the physical reason at the core of this theoretical result.


\appendix
\section{Photon self-energy eigenvalues}\label{app}

	

	
The eigenvalues of the  photon self-energy  in vacuum were calculated using one-loop approximation in \citet{Shabad1975}:
	\begin{equation}\label{PolOp1}
	\kappa^{(i)}  =\dfrac{2 \alpha}{\pi}\int_{0}^{\infty}d\tau\int_{-1}^{1}d\eta    e^{-\dfrac{\tau}{b}}   [   \rho_{i} e^{\zeta}+\dfrac{(k^2_\parallel+k_{\perp}^2-\omega^2)\bar{\eta}^2}{2t}  ],
	\end{equation}
	
\begin{eqnarray*}
		\zeta&=&\frac{1}{bm^2} \left[ (\omega^2-k^2_{\parallel})\bar{\eta}^2\tau-k^2_{\perp}\frac{\sinh{(\eta_+\tau)}\sinh{(\eta_-\tau)} }{\sinh{\tau}} \right],\\
	\rho_1 &=&-\dfrac{ k^2 }{2}\frac{\sinh{(\eta_+\tau)}\cosh{(\eta_+\tau)} }{\sinh^2{\tau}}\eta_-, \\
	\rho_2&=&\frac{\omega^2-k^2_{\parallel}}{2}\bar\eta^2 \coth{\tau}-\dfrac{k^2_{\perp}}{2}\frac{\sinh{(\eta_+\tau)}\cosh{(\eta_+\tau)} }{\sinh^2{\tau}}\eta_-,\\
	\rho_3&=&\dfrac{\omega^2-k^2_{\parallel}}{2}\dfrac{\sinh{(\eta_+\tau)}\cosh{(\eta_+\tau)} }{\sinh^2{\tau}}\eta_--\dfrac{k^2_{\perp}}{2}\frac{\sinh{(\eta_+\tau)}\sinh{(\eta_-\tau)} }{\sinh^3{\tau}}.
\end{eqnarray*}
where 
	$\alpha $ is the fine-structure constant, $\eta_{\pm}=\frac{1 \pm \eta}{2}$, $\bar{\eta}=\sqrt{\eta_+\eta_-}$.
	

We start by rewriting the functions $\zeta, \rho_i$ in terms of $k^2$ and $k_\parallel^2-\omega^2$, by making the substitution  $k_\perp^2=k^2-(k_\parallel^2-\omega^2) $ in Eq.~(\ref{PolOp1}). Then, we use the series expansion of the exponential $e^{\zeta}$ around the light-cone $\frac{k^2}{eB}=0$, to get
\begin{equation}\label{PolOpSeries}
    \kappa_{i}=\sum_{l=0}^{\infty}\chi_{il}(k^2)^l,
		\end{equation}
where the the coefficients $\chi_{il}$depend on the magnetic field $b$ and $k_\parallel^2-\omega^2$:
	\begin{eqnarray} \label{chi}
	\chi_{il}&=&\dfrac{2 \alpha}{\pi}\int_{0}^{\infty}d\tau\int_{-1}^{1}d\eta \,  e^{-\dfrac{\tau}{b}} [ e^{\zeta_0} \left (\dfrac{\xi}{b m_{e}^2}\right )^{l} \nonumber \\  & & \hspace{.8cm} \dfrac{(-1)^{l}}{l!} \left( \rho_{0i}-l\left (\dfrac{\xi}{b m_{e}^2}\right )^{-1} \theta_{i} \right)+\delta_{1l}\dfrac{\bar{\eta}^2}{2\tau}
	],
	\end{eqnarray}
and
	
	
%
\begin{eqnarray*}
	\zeta_{0}&=&\frac{(\omega^2-k^2_{\parallel})(\bar{\eta}^2 \tau-\xi)}{b m_{e}^2},  	
	\\
	\rho_{01} &=& 0, \\
	\rho_{02}&=&\dfrac{\omega^2-k^2_{\parallel}}{2}(\bar{\eta}^2 \coth{\tau}-\varrho ),\\
	\rho_{03}&=&\dfrac{\omega^2-k^2_{\parallel}}{2}(\varrho-\dfrac{\xi}{\sinh^2{\tau}} ), \\
	\theta_{1}&=&\theta_{2}=-\dfrac{\varrho}{2},\hspace{1cm}
	\theta_{3}=-\dfrac{\xi}{2 \sinh^2{\tau}},
	\end{eqnarray*}
and we have also used the auxiliary functions	
		\begin{equation}
	   \xi= \frac{\sinh{(\eta_+\tau)}\cosh{(\eta_+\tau)} }{\sinh{\tau}},\quad
	   \varrho=\frac{\sinh{(\eta_+\tau)}\cosh{(\eta_+\tau)} }{\sinh^2{\tau}}.
	   \end{equation}
	
\bibliography{references}%



\end{document}